\documentclass[fleqn,usenatbib]{mnras}

\usepackage{newtxtext,newtxmath}
\usepackage[T1]{fontenc}
\usepackage{blindtext}
\usepackage{hyperref}

\DeclareRobustCommand{\VAN}[3]{#2}
\let\VANthebibliography\thebibliography
\def\thebibliography{\DeclareRobustCommand{\VAN}[3]{##3}\VANthebibliography}

\usepackage{graphicx}	% Including figure files

\usepackage{amsmath}	% Advanced maths commands
\usepackage{multirow}

\newcommand{\dd}{{\rm d}}
\newcommand{\msun}{\mbox{M$_\odot$}}
\newcommand{\myr}{\mbox{${\rm Myr}$}}
\newcommand{\gyr}{\mbox{${\rm Gyr}$}}
\newcommand{\pc}{\mbox{${\rm pc}$}}
\newcommand{\cpc}{\mbox{${\rm cpc}$}}

\newcommand{\rh}{{r_{\rm h}}}
\newcommand{\msunpc}{{\msun\,\pc^{-3}}}

\newcommand*\code[1]{\textsc{#1}}

\newcommand{\emppathfinder}{\textsc{EMP}-\textit{Pathfinder}\xspace}

\newcommand{\arepo}{\textsc{arepo}\xspace}

\newcommand{\grackle}{\code{Grackle}\xspace}

\defcitealias{Reina-Campos22}{Paper I}
\title[Cluster sizes and dissolution]{Initial sizes of star clusters: implications for cluster dissolution during galaxy evolution}

\author[M. Reina-Campos et al.]{
Marta Reina-Campos,$^{1,2}$\thanks{E-mail: reinacampos@mcmaster.ca}
Alison Sills,$^{1}$
Godefroy Bichon,$^{3,1}$
\\
$^{1}$Department of Physics \& Astronomy, McMaster University, 1280 Main Street West, Hamilton, L8S 4M1, Canada\\
$^{2}$Canadian Institute for Theoretical Astrophysics (CITA), University of Toronto, 60 St George St, Toronto, M5S 3H8, Canada\\
$^{3}$ École Supérieure de Physique et de Chimie Industrielles de la Ville de Paris (ESPCI), 10 Rue Vauquelin, 75005 Paris, France\\
}

\date{Accepted XXX. Received YYY; in original form ZZZ}

\pubyear{2023}

\begin{document}
\label{firstpage}
\pagerange{\pageref{firstpage}--\pageref{lastpage}}
\maketitle

% Abstract of the paper
\begin{abstract}
Massive star clusters are often used as tracers of galaxy formation and assembly. In order to do so, we must understand their properties at formation, and how those properties change with time, galactic environment, and galaxy assembly history. The two most important intrinsic properties that govern star cluster evolution are mass and radius. In this paper, we investigate 10 theoretically and observationally motivated initial size-mass relations for star clusters, and evolve populations of clusters through galaxy formation models. We compare our results to each other and to observations of cluster populations in M83, M31, and the Milky Way. We find that none of our size-mass relations agree with the observations after 6-10 Gyr of evolution. We can successfully reproduce the cluster mass functions with models that have a small range of initial radii, and which do not allow cluster radii to change with time. However, these models do not agree with our understanding of cluster evolution, which does involve radius evolution, and do not match the observed distributions of radii. We note that there is a region of parameter space where clusters are optimally protected from both tidal shocks and evaporation due to two-body relaxation. Clusters which are allowed to evolve into this parameter space will likely survive. An improved understanding of both mass and radius evolution of star clusters in realistic, time-varying galactic potentials is necessary to appropriately make the connection between present-day cluster properties and their use as tracers of galaxy formation and assembly.

\end{abstract}


%%%%%%%%%%%%%%%%%%%%%%%%%%%%%%%%%%%%%%%%%%%%%%%%%%

%%%%%%%%%%%%%%%%% BODY OF PAPER %%%%%%%%%%%%%%%%%%

\section{Introduction}

Globular clusters (GCs) -- old, massive, metal-poor, bound stellar systems -- are found in galaxies of all types. They are compact and luminous, so they are relatively easy to identify and characterize at large distances. They are often treated as simple stellar systems with (approximately) a single age and metallicity. All of these qualities taken together mean that globular clusters, and their younger counterparts Young Massive Clusters (YMCs), are often used as probes of galaxy formation and assembly. The process of galaxy assembly involves mergers of galaxies, and those galaxies bring their own cluster systems with them. The properties of a galaxy's globular cluster system, such as the distribution of metallicities, ages, and masses, provide clues about the environment in which those clusters formed. 

In order to successfully use star clusters to probe galaxy evolution, we must understand both the initial properties of clusters, and their subsequent evolution. Clusters are formed in dense gas, likely in high-pressure environments \citep{Elmegreen1997}. They inherit the chemical composition of the gas out of which they are formed. The gas properties determine the initial mass and size\footnote{We use `half-mass radius' and `size' interchangeably in this work.} of the cluster \citep[e.g.][]{ReinaCampos2017,TrujilloGomez2019}, and where in the galaxy they will form. For this aspect, we need an appropriate treatment of the evolution of the interstellar medium (ISM) in galaxies and an understanding of how the ISM properties determine the initial cluster properties. 

In addition, we must properly model all the processes which can modify the initial cluster properties. Mass is lost from clusters because of stellar evolution, two-body interactions in the cluster and stripping in the galactic tidal field \citep{Lamers2010}, from tidal shocks with nearby giant molecular clouds \citep{Gieles16} or the galactic disk \citep{webb2014}, and through dynamical friction \citep{Fellhauer2007}. If these processes are efficient enough, the cluster will be fully disrupted (i.e. dissolved). The present-day population of clusters are only those that have survived as massive bound objects. The radius of the cluster is likely to change as the cluster loses mass. To follow the mass and radius evolution, we require knowledge of both the internal dynamics of the cluster, and the impact of the cluster's local environment such as its location in the galaxy and hence the tidal effects of the galaxy on the cluster. 

Finally, we must also have an understanding of galaxy assembly. Clusters may have formed in the main galaxy progenitor, or they may have formed in a satellite galaxy which subsequently merged with the main galaxy, or they may have formed during a gas-rich merger. Those cluster populations will likely have different properties, and therefore we can use those properties to determine how many, and what kinds, of events went into the creation of the current galaxy. Cosmological-scale simulations such as EAGLE \citep[Evolution and Assembly of GaLaxies and their Environments;][]{Schaye15, Crain15} are often used for this aspect of the problem.  

In recent years, the improvements in both computing power and our understanding of the important physical processes have allowed a number of groups to approach the problem of cluster formation and evolution using multi-scale, multi-physics simulations \citep{Li17,Li18,Pfeffer18,Kruijssen19a,Brown22,Reina-Campos22}. These simulations include a cosmological-scale treatment of dark matter; a model of the interstellar medium including heating and cooling, enrichment, feedback, and star formation; and treatment of the interplay between galaxy environment and cluster formation, evolution, and disruption. These simulations have been very successful in predicting overall properties of globular cluster systems in present-day galaxies, and in disentangling sub-populations of clusters in the Milky Way \citep[e.g.][]{Kruijssen19b,Kruijssen20}.

We are not yet at the point where a single simulation can simultaneously treat all scales and all physical processes at the necessary level of detail. Therefore, all of these simulations have some mass and/or spatial resolution scale below which physical processes are treated with a sub-grid model. In some cases \citep[e.g.][]{Choksi18} the model is explicitly given as scaling relations of quantities with host galaxy mass; in other cases \citep[e.g.~E-MOSAICS;][]{Pfeffer18,Kruijssen19a} the local galactic environment is used to determine the properties of clusters as they form and evolve. These sub-grid models are necessarily simplistic, and are often created from analytic prescriptions for physical processes that depend on global or averaged properties. For example, the clusters with different concentrations respond differently to a tidal field, but in sub-grid models, a typical concentration is assumed. The tidal field is modelled on scales larger than the size of a typical star cluster, so its effect at smaller scales must be estimated.  Nevertheless, despite their simplifications, models of cluster evolution during galaxy formation and assembly predict the observed properties of cluster populations remarkably well, and therefore have become a standard component of galaxy assembly models.

Sub-grid models of cluster evolution necessarily include assumptions about the initial mass functions and initial radii of stellar clusters. The initial cluster mass function (ICMF) has received a lot of attention, both observationally and theoretically \citep[for a review, see][]{Adamo2020}. Masses of clusters are fairly straightforward to determine observationally, and so the present-day mass functions of clusters of a variety of ages are commonly used as tests of cluster formation and evolution models. Cluster radii, on the other hand, are harder to measure, particularly when the clusters are in distant galaxies and therefore unresolved. Our Milky Way is not currently forming very massive clusters in large numbers, so we must turn to the small sample of sufficiently nearby star-forming galaxies to learn about about the initial cluster radius distribution \citep[][]{Ryon15,BrownGnedin2021}. In addition, there are estimates of the radii of proto-globular cluster candidates at high redshift in highly lensed fields \citep[e.g.][]{Vanzella22a,Vanzella22b,Claeyssens2022} which make use of the first results from JWST. We should have direct measurements of the initial properties of clusters over a large range of formation redshifts available within the next few years, but we are not yet at that stage. Most sub-grid models for cluster formation therefore use simple radii distributions such as assuming that all clusters are born with the same radius, or that all clusters are born with the same density. Recently, a few groups are starting to investigate an environmentally-dependent initial radius distribution \citep[][]{ChoksiKruijssen2021,Webb21}.

It is important for sub-grid models of cluster evolution that the treatment of cluster radii is realistic. Some of the cluster disruption mechanisms that are a necessary component of the models depend on the cluster radius or the cluster density. For example, the mass loss rate due to two-body relaxation and subsequent tidal stripping in the galactic tidal field is governed by the size of the cluster relative to its tidal radius. The mass loss rate due to tidal shocks is proportional to the half-mass radius of the cluster and the strength of the shock. In this paper, we undertake a systematic exploration of the treatment of cluster radii within a galaxy formation model. Specifically, we are using the \emppathfinder galaxy formation model \citep[][Paper I]{Reina-Campos22}. These cosmological zoom-in simulations of Milky Way-mass galaxies include a
multi-phase ISM and a simultaneous self-consistent star cluster formation and evolution model that is an updated version of MOSAICS \citep[MOdelling Star cluster population Assembly In Cosmological Simulations,][]{Kruijssen11,Pfeffer18}. For this exploration of initial radii, we select one galaxy from the \emppathfinder suite, and implement 10 different radius distributions drawn from suggestions in the literature. Specifically, we explore the effects of all clusters having the same initial radius, all clusters having the same initial density, a radius distribution that depends on the cluster formation environment, and empirical distributions based on observations of young clusters.  Additionally, we explore the effect of size evolution on the final cluster populations by re-running the same galaxy and allowing the cluster sizes to evolve over cosmic time. The sub-grid framework implemented in \emppathfinder allows us to compare the different assumptions of the initial properties of stellar cluster populations, as well as their subsequent evolution, within the same galactic environment.

The paper is structured as follows. In Section \ref{sec:methods}, we describe the galaxy formation model, how the clusters' masses and radii can evolve, and our choices for the initial radius distributions. In Section \ref{sec:results}, we describe the results of our models and compare the resulting cluster populations to those in M83, M31 and the Milky Way. We summarize our results, and discuss their implications in Section \ref{sec:conclusions}.

%Methods and models
\section{Methods}\label{sec:methods}

In this section we briefly describe the \emppathfinder galaxy formation model (Sect.~\ref{sub:emppathfinder}), with an emphasis on how clusters lose mass, and we discuss the initial size-mass relations considered (Sect.~\ref{sub:models-initial-rh}) and how cluster sizes evolve with time (Sect.~\ref{sub:models-size-evolution}).   

\subsection{The galaxy formation model: \emppathfinder}\label{sub:emppathfinder}

We use the \emppathfinder galaxy formation model \citepalias{Reina-Campos22} to study the formation and evolution of sub-grid stellar cluster populations in Milky Way-mass galaxies over cosmic time. The baryonic physics of this model are implemented in the moving-mesh hydrodynamical code \arepo \citep{Springel10}. In this work, we use the initial conditions corresponding to the Milky Way-mass galaxy MW$04$ from \citetalias{Reina-Campos22}, and we run them using the \emppathfinder galaxy formation model. We define the baryonic target mass to be $m_{\rm gas} = 2.26\times10^5~\msun$, and the smallest gravitational softening for the gas cells is $\epsilon_{\rm gas} = 80~\cpc$.

A key ingredient of the model are the physics of the multi-phase nature of the interstellar medium (ISM), which enables studies of how the cold, gas reservoir in galaxies affects the formation and evolution of star clusters. For this, we model the thermodynamic state of the gas using the simplest non-equilibrium network from the chemistry library \grackle \citep{Smith17}. We use the six species network, which solves for the chemical abundances, heating and radiative cooling for H, H$^{+}$, He, He$^{+}$ and He$^{++}$. This network does not assume ionization equilibrium for the species, but rather solves for their abundances instead. In the fiducial model, we use a constant star formation efficiency to calculate the star formation rate of gas cells once they become sufficiently cold ($T<1.5\times10^4~\rm K$) and dense ($n>1~\rm H~cm^{-3}$). The star formation efficiency per free-fall time is assumed to be $20~$per cent, and the conversion of gas cells into stars is determined in a probabilistic basis. As the stellar particles evolve, we follow the mass, metals, yields, energy and momentum ejected by supernovae of type II and Ia, as well as by winds from the asymptotic giant branch. For additional details regarding the baryonic physics included, we refer the reader to \citetalias{Reina-Campos22}.

Whenever a gas cell is converted into a star particle, sub-grid stellar clusters are spawned. We use the framework implemented within \emppathfinder to model 10 different cluster populations within a single Milky Way-mass galaxy over cosmic time. For each of these cluster populations, we assume a different initial size-mass relation while leaving all the other models as in the `Fiducial' population from \citetalias{Reina-Campos22} (see Table~\ref{tab:sum-input-models-clusters}). This model reproduces the observed properties of the globular cluster populations of the Milky Way and M31. 

\begin{table*}
\centering{
  \caption{Summary of the input models used to model the sub-grid stellar cluster populations. Except for the assumptions regarding the initial sizes and their evolution, we use the same combination of input models as in the `Fiducial' model from \citetalias{Reina-Campos22}. In addition to tidal shocks, the sub-grid stellar clusters also lose mass due to stellar evolution, two-body relaxation and dynamical friction.}
  \label{tab:sum-input-models-clusters}
	\begin{tabular}{ccccc}\hline
		Cluster formation efficiency & Initial cluster mass function & Initial half-mass radius & Tidal shock disruption & Size evolution \\ \hline
		$\Gamma(\Sigma_{\rm gas}, \kappa, Q)$ & Double Schechter function & See Sect.~\ref{sub:models-initial-rh} & $t_{\rm sh} \propto \rho \left[\sum_{ij}\left(\int T_{ij} dt\right)^2 A_{{\rm w}, ij}\right]^{-1}$ & See Sect.~\ref{sub:models-size-evolution} \\
		& $M_{\rm cl, min}(\Sigma_{\rm gas}, \kappa, Q)$ \& $M_{\rm cl, max}(\Sigma_{\rm gas}, \kappa, Q)$ &  &  &  \\
		\citet{Kruijssen12} & \citet{TrujilloGomez2019} & -- & \citet{Kruijssen11} & -- \\
		\hline 
	\end{tabular}}
\end{table*}

The number of clusters, as well as their masses, are determined by the gas conditions around the newborn star particle. We use an environmentally-dependent model for the cluster formation efficiency from \citet{Kruijssen12}, and we assume that the initial masses of clusters are well described by an environmentally-dependent double Schechter function, i.e.  a power-law of slope $\alpha=-2$ with low- and high-mass truncation exponential cut-offs \citep{ReinaCampos2017,TrujilloGomez2019}. In order to explore the influence of the initial size-mass relation on the resulting stellar cluster populations, we consider ten initial size-mass relations suggested in the literature. We discuss them in more detail in Sect.~\ref{sub:models-initial-rh}.

After clusters have been formed, we follow the evolution of their masses over time considering different disruption mechanisms. Firstly, stellar clusters lose mass due to stellar evolution. Secondly, we account for dynamical disruption due to tidal shocks and two-body relaxation. Dynamical mass loss is governed by the local tidal field, and by the masses and sizes of clusters. We follow the variation of the local tidal field at the location of clusters during run-time to determine the amount of mass lost. If a quick variation is identified\footnote{Any quick perturbation of the local tidal tensor is assumed to inject energy to the cluster, i.e. we do not distinguish between whether the perturbation is caused by giant and dense molecular clouds or crossing of the galactic disk.}, we calculate the mass loss due to the corresponding tidal heating that is injected in the cluster, and reduce the mass of the cluster accordingly \citepalias[see sect.~3.2.1 from][]{Reina-Campos22}. Once the mass of a cluster is smaller than $100~\msun$, we consider that it has been completely disrupted as it is no longer dense enough to remain gravitationally bound.

Tidal shocks with passing dense and cold clouds are expected to dominate cluster disruption, both at the early stages of their evolution and over their entire evolution during a Hubble time. We model shock-driven mass loss as inversely proportional to the shock timescale, $(\dd m/\dd t)_{\rm sh} = - m/t_{\rm sh}$, which describes the disruption time for a given cluster \citep{Kruijssen11},
\begin{equation}
t_{\rm sh} \propto m\,\rh^{-3} \left[\sum_{ij}\left(\int T_{ij} dt\right)^2 A_{{\rm w}, ij}\right]^{-1}.
\end{equation}
The shock timescale depends on the mass $m$ and half-mass radius $\rh$ of the cluster, and on the components of the local tidal field, $T_{ij}$. The Weinberg correction factors, $A_{{\rm w}, ij}$, describe the amount of injected energy absorbed by the adiabatic expansion of the cluster \citep{Weinberg94a,Weinberg94b,Weinberg94c,Gnedin03}, such that the energy injected in dense clusters expands them rather than disrupts them. For a cluster of a fixed mass and given a tidal shock, more extended stellar clusters would require less time to get disrupted.

Similarly, we consider the continuous mass loss driven by two-body interactions within the cluster, $(\dd m/\dd t)_{\rm rlx} = - \xi m/t_{\rm rlx}$. We describe it in terms of the relaxation timescale at the half-mass radius \citep{Spitzer71,Giersz94a}, $t_{\rm rlx} \propto {m^{1/2} \rh^{3/2}}{\ln\left(\gamma m/\bar{M}\right)}^{-1}$, where $\gamma \approx 0.11$ for equal-mass clusters \citep{Giersz94a} and $\bar{M} = 0.42~\msun$ is the mean stellar mass in a \citet{Chabrier05} IMF integrated between $0.08\leq m \leq 120~\msun$. For a given tidal environment, more extended or more massive clusters require longer to disrupt due to two-body evaporation alone.

Hence, dynamical disruption mechanisms are expected to affect stellar clusters differently depending on their mass and size. For a given mass, if a cluster is more extended, tidal shocks can inject more energy to stars closer to the perturber rather than to those farther away, thus creating the tidal effect. At the same time, the larger size avoids two-body interactions between the stars. On the contrary, if the cluster is more compact, interactions between the star are more likely due to their smaller distances, but tidal effects are weaker because the energy injection is homogeneously distributed across the cluster.  Therefore, for a given mass, clusters with a larger radius are more susceptible to be disrupted by tidal shocks, whereas those with smaller radius are more sensitive to two-body evaporation.

The last disruption mechanism that we consider is the effect of dynamical friction. We have to do so in post-processing due to the sub-grid nature of the stellar clusters. To account for this effect, we completely disrupt those clusters whose age is less than the timescale required to in-spiral to the center of their host galaxy. 

Lastly, the half-mass radius of the sub-grid clusters in \emppathfinder can evolve over time. Clusters can expand adiabatically due to stellar evolution and evaporation, and they can either expand or contract due to tidal shocks. We discuss further the physics of size evolution and our choices in Sect.~\ref{sub:models-size-evolution}.

\subsection{Initial mass-radii relations}\label{sub:models-initial-rh}

\begin{table*}
\centering{
  \caption{Summary of the initial size-mass relations assumed in this work. From left to right, columns contain the name of the model, the initial size-mass relation, the scatter in natural logarithm of the size, and the reference. In the models with scatter, for each cluster we sample their size from a log-normal distribution centered on their initial size given their mass, $\ln[\rh^{\rm init}(m)]$, and with scatter $\sigma_{\ln \rh}$.}
  \label{tab:sum-initial-size-mass}
	\begin{tabular}{lccc}\hline
		Model & Initial size-mass relation & Scatter in $\ln(\rh)$ & Reference\\ \hline
		Constant radius & $\rh^{\rm init} = 4~\pc$ & -- & -- \\
		Physical conditions & $\rh^{\rm init} \propto m^{1/2} \Sigma_{\rm gas}^{-1/2}$% %= f(m, \Sigma_{\rm gas})$ 
        & $0.3~{\rm dex}$ & \citet{ChoksiKruijssen2021} \\
		Constant density & \multirow{2}{*}{$\rh^{\rm init} =  3~\pc\left(m/10^4~\msun\right)^{0.3}$} & -- & -- \\
		Constant density with scatter & & $0.3~{\rm dex}$ & -- \\ \hline
		Fit to $1$--$10~\myr$ & \multirow{2}{*}{$\rh^{\rm init} = 2.365~\pc \left(m/10^4~\msun\right)^{0.180}$} & -- & \multirow{2}{*}{table 2 from \citet{BrownGnedin2021}} \\
		Fit to $1$--$10~\myr$ and scatter & & $0.735~{\rm dex}$ & \\
		2D distribution for $1$--$10~\myr$ & $\rh^{\rm init} = f(m)$ & $\sigma = f(m)$ & \citet{BrownGnedin2021} \\
		Fit to $10$--$100~\myr$ & \multirow{2}{*}{$\rh^{\rm init} = 2.506~\pc \left(m/10^4~\msun\right)^{0.279}$} & -- & \multirow{2}{*}{table 2 from \citet{BrownGnedin2021}} \\
		Fit to $10$--$100~\myr$ and scatter &  & $0.548~{\rm dex}$ & \\
		2D distribution for $10$--$100~\myr$ & $\rh^{\rm init} = f(m)$ & $\sigma = f(m)$ & \citet{BrownGnedin2021} \\
		\hline 
	\end{tabular}}
\end{table*}

The aim of this paper is to explore the effect of different initial size-mass relations on the evolution of stellar clusters over a cosmic time. To do this, we consider ten different relations that are suggested in the literature, and that we summarize in Table~\ref{tab:sum-initial-size-mass}. For the models with scatter, for each cluster we sample their size from a log-normal distribution centered on their initial size given their mass, $\ln[\rh^{\rm init}(m)]$, and with scatter $\sigma_{\ln \rh}$. In other words, these are the parameters of the log-normal function itself, not the underlying normal distribution.

The first four initial size-mass relations that we consider are theoretically-motivated:
\begin{itemize}
    \item \emph{Constant radius}: all stellar clusters are given a constant radius of $\rh^{\rm init}=4~\pc$. This corresponds to the same choice as in the `Fiducial' population used in \citetalias{Reina-Campos22}.
    \item \emph{Physical conditions}: the initial half-mass radius is given by the natal gas conditions of stellar clusters, $\rh^{\rm init} \bf{\propto m^{1/2} \Sigma_{\rm gas}^{-1/2}}$%= f(m, \Sigma_{\rm gas})$ 
    , according to \citet{ChoksiKruijssen2021}.
    \item \emph{Constant density}: the initial half-mass radius is set by the mass of the cluster, $\rh^{\rm init} \propto m^{0.3}$, so that all clusters have close to the same density. 
    \item \emph{Constant density with scatter}: Same relationship as above, but assuming a scatter of $0.3~\rm dex$ in $\ln(\rh)$ at a constant mass. 
\end{itemize}

The next six initial size-mass relations are empirical. They are derived from the catalogue of effective sizes of the LEGUS\footnote{Legacy ExtraGalactic UV Survey with The Hubble Space Telescope - \href{https://legus.stsci.edu/legus_survey.html}{https://legus.stsci.edu/legus\_survey.html} \citep{Calzetti15,Adamo17}} star clusters compiled by \citet{BrownGnedin2021}, under the assumption that the effective radius is the half-mass radius of the cluster:
\begin{itemize}
    \item \emph{Fit to $1$--$10~\myr$}: the initial half-mass radius of clusters is given by a linear relationship ($\rh \propto m^{0.180}$) that flattens to a constant value at the high-mass end ($m\geq10^5~\msun$). This relationship corresponds to the linear fit to the clusters in the $1$--$10~\myr$ age range as provided in table 2 from \citet{BrownGnedin2021}.
    
    \item \emph{Fit to $1$--$10~\myr$ with scatter}: we use the same relationship as above, including a scatter of $0.735~\rm dex$ in $\ln(\rh)$ at a constant mass. This scatter corresponds to the intrinsic scatter quoted in \citet{BrownGnedin2021} for this age range.
   
    \item \emph{2D distribution for $1$--$10~\myr$}: the initial half-mass radius is determined from a binned distribution. To create the distribution, we binned the empirical effective radius of clusters $1$--$10~\myr$ old from \citet{BrownGnedin2021} between $10^2$--$10^6~\msun$. Within each bin, we then determine the mean and the standard deviation in the sizes in logarithmic space, which we use at run-time to generate the initial half-mass radii. When a bin has relatively few members with a fair number of outliers, for example at approximately 10$^5~\msun$, this can result in unphysical structure in our initial mass-radius relation. For model clusters above $10^6~\msun$, we use a constant radius with a scatter equal to the scatter at 10$^6~\msun$.
    \item \emph{Fit to $10$--$100~\myr$}: the initial half-mass radius of clusters is given by a linear relationship ($\rh \propto m^{0.279}$) that flattens to a constant value at the high-mass end ($m\geq10^5~\msun$). This relationship corresponds to the linear fit to the clusters in the $10$--$100~\myr$ age range as provided in table 2 from \citet{BrownGnedin2021}.
    
    \item \emph{Fit to $10$--$100~\myr$ with scatter}: we use the same relationship as above, including a scatter of $0.548~\rm dex$ in $\ln(\rh)$ at a constant mass. This scatter corresponds to the intrinsic scatter quoted in \citet{BrownGnedin2021} for this age range.
    
    \item \emph{2D distribution for $10$--$100~\myr$}: the initial half-mass radii are given by a binned distribution. We create the distribution similarly to the description above, but using the sample of clusters $10$--$100~\myr$ old from \citet{BrownGnedin2021}.
\end{itemize}

\begin{figure*}
	\includegraphics[width=\hsize,keepaspectratio]{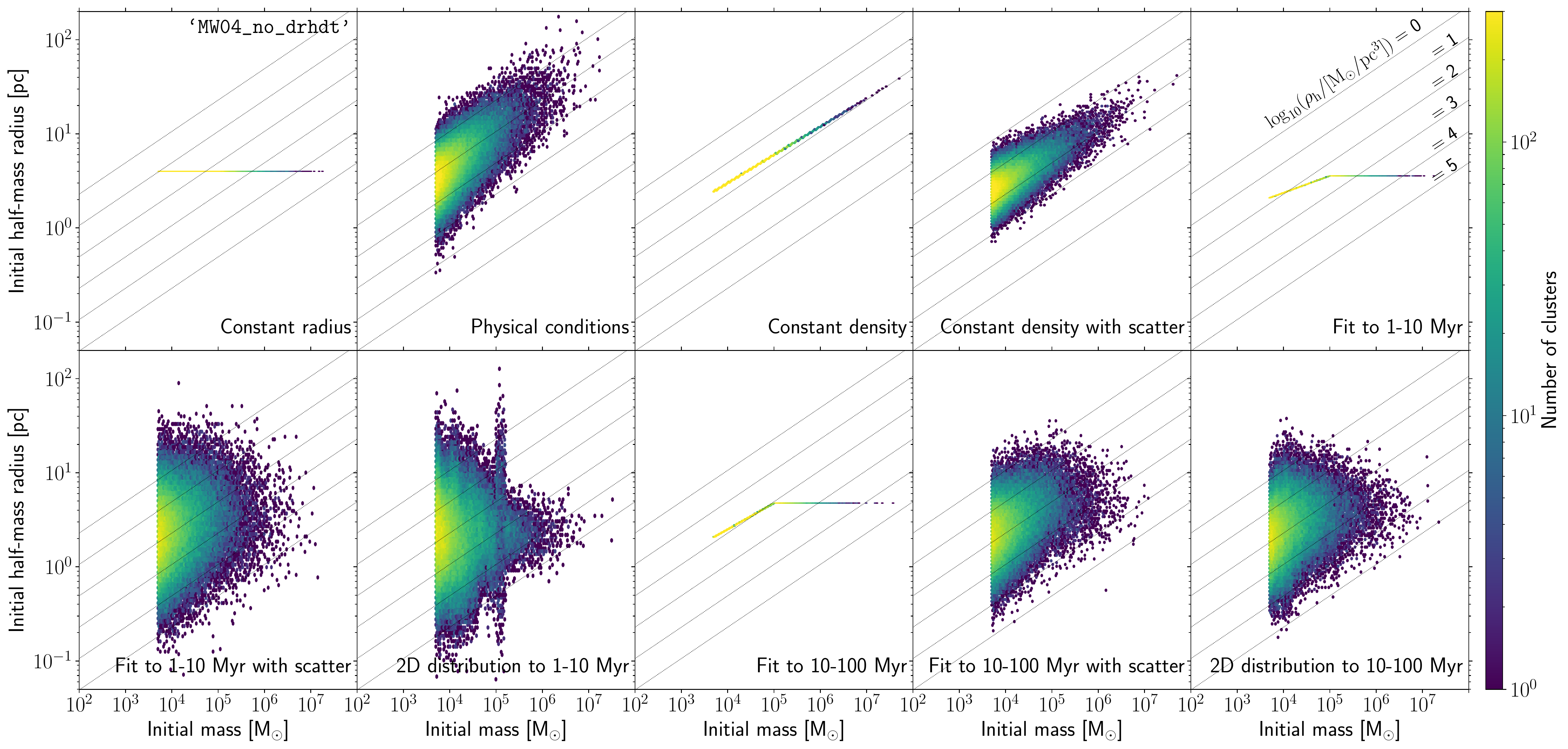}
    \caption{Initial cluster half-mass radius as a function of their initial mass for our 10 sub-grid cluster populations. The colorbar shows the number of clusters per bin, and it ranges between $1$--$300$ in logarithmic space. Diagonal dotted lines correspond to lines of constant density.}
    \label{fig:mass_rad_init}
\end{figure*}

We show the initial sizes (half-mass radii) and masses of stellar clusters for our 10 populations in Fig.~\ref{fig:mass_rad_init}. Clusters in our models are limited to have masses greater than $5 \times 10^3~\msun$.

For comparison, the proto-globular cluster candidates found in lensed high-redshift fields \citep{Vanzella22a,Vanzella22b} have inferred masses of $\sim10^5$--$10^6~\msun$ and sizes of a few to tens of parsecs. The vast majority of clusters in our populations have densities between $10$--$10^3~\msunpc$, but there is variety among our ten initial size-mass relations. For the relations without scatter, clusters can have densities larger than $10^4~\msunpc$ only at very large masses ($m\gg10^6~\msun$).
In contrast, those clusters sampled from the empirical relations with scatter show a large spread to both high ($\rho > 10^4~\msunpc$) and low ($\rho < 1~\msunpc$) densities across the mass range. The populations with the widest spread in densities are `Fit to $1$--$10~\myr$ with scatter' and `2D distribution to $1$--$10~\myr$'. The former forms $2$~per cent of the clusters with initial densities below $\rho < 1~\msunpc$ and $2.2$~per cent of them with densities larger than $\rho > 10^4~\msunpc$, whereas the latter forms $2.5$~per cent and $4.6$~per cent, respectively, for the same density cuts.

\subsection{Size evolution}\label{sub:models-size-evolution}

In addition to the initial cluster size-mass relation, whether or not sizes change over time can also have an effect on the evolution of stellar clusters over cosmic time. To investigate this, we consider two different assumptions regarding the size evolution of stellar clusters.

In the first set of models (\texttt{`MW04\_no\_drhdt'}), we assume there is no evolution of the cluster radius, even as the clusters lose mass. This assumption allows us to identify the relation between the \emph{initial} cluster properties and cluster disruption. This assumption is also used in the `Fiducial' cluster population from \citetalias{Reina-Campos22}.

In the second set of models (\texttt{`MW04\_drhdt'}), we allow the cluster half-mass radius to evolve with time. The evolution of the radius responds to the mass lost via stellar evolution, evaporation and tidal shocks \citepalias[see sect.~3.2.2. in][]{Reina-Campos22}. The half-mass radii can expand adiabatically due to mass lost during stellar evolution. The amount of expansion is given by the ratio of the stellar particle mass at the previous timestep relative to the current one. As when applying the mass evolution, this term is applied after accounting for the effect of the dynamical processes. 

We describe the role of dynamical processes in the evolution of the size of star clusters with two terms,
\begin{equation}
    \left(\dfrac{\dd \rh}{\rh}\right)_{\rm dyn} = \left( 2 - \dfrac{1}{f}\right)\left(\dfrac{\dd m_{\rm sh}}{m}\right) +  \left( 2 - \dfrac{\zeta}{\xi}\right)\left(\dfrac{\dd m_{\rm rlx}}{m}\right)
    \label{eq:size-evolution-complete}
\end{equation}
that are linked to the influence of tidal shocks and evaporation in driving cluster disruption, $\left(\dd m_{\rm sh}/m\right)$ and $\left(\dd m_{\rm rlx}/m\right)$, respectively. The terms $f$, $\zeta$ and $\xi$ correspond to the fraction of the relative energy change that is converted to a change in cluster mass due to shocks, to a coefficient that relates the relaxation-driven energy evolution to the relaxation timescale, and to the fraction of escaper stars per relaxation time, respectively.

At a given moment in time, if mass loss from shocks dominates over that of evaporation ($\Delta m_{\rm sh} > \Delta m_{\rm rlx}$), the cluster size evolution is dominated by the first term of equation (\ref{eq:size-evolution-complete}), which causes clusters to contract. In contrast, if evaporation dominates over shock-driven disruption, then the size evolution is dominated by the second term of equation (\ref{eq:size-evolution-complete}), which leads to an overall expansion of the cluster. The size evolution caused by stellar evolution and dynamical processes can lead to size-mass relations that are dramatically different from the initial relation.

\subsection{Observational data: Milky Way, M31 and M83}

We compare our models to three different cluster populations: those in the Milky Way, M31 and M83. For the Milky Way, we use the catalogue by \cite{Baumgardt18}, which provide the masses and structural parameters for 112 globular clusters. In the case of M31, we use the masses from the study by \cite{Caldwell16}, and we combine those with the half-mass radii provided by \cite{Barmby07} and \cite{Huxor14}. Thus, we have estimates for the ages and masses of 439 clusters, which reduce to 79 objects with age, mass and size estimates. Lastly, \cite{Ryon15} provides the masses and sizes of the young star clusters in M83.

\section{Results}\label{sec:results}

In this section we present the evolved size-mass relations of the ten sub-grid populations, and we explore which physical mechanism dominates the evolution of sizes over time. We remind the reader that the total number of clusters formed in each of our 10 models, and their masses, are the same between all of the models presented here as we have fixed the underlying galaxy assembly history by choosing only one galaxy from \citetalias{Reina-Campos22} and by using the same initial cluster mass function prescription. Therefore, differences in the total number of clusters or in their radii and mass distributions between models is a result of the initial radius distribution, and size evolution if it is included.

\subsection{Final size-mass distributions of cluster populations}\label{sub:final-size-mass-relations}

\begin{figure*}
	\includegraphics[width=\hsize,keepaspectratio]{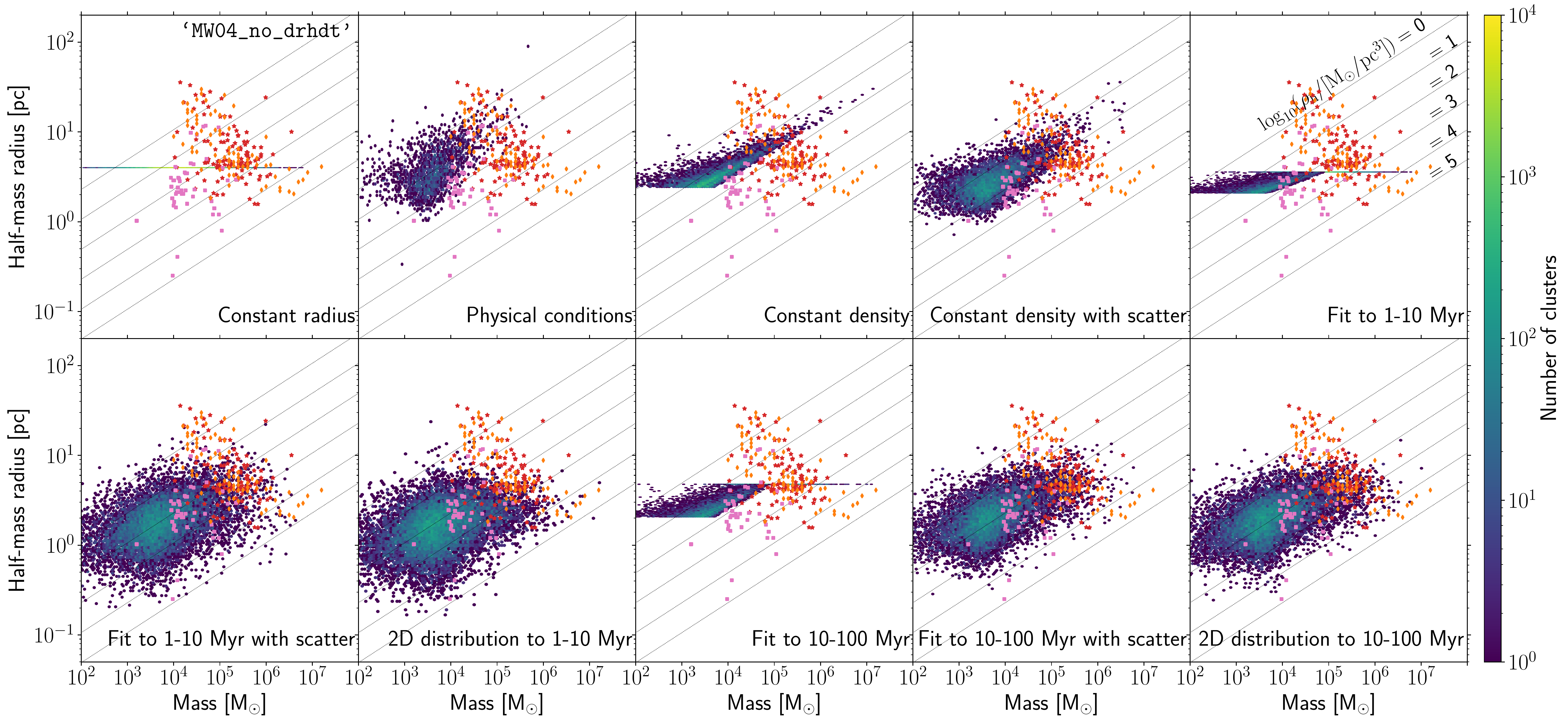}
    \caption{Cluster half-mass radius as a function of their final masses for our 10 sub-grid cluster populations in the Milky Way-mass galaxy without size evolution (\texttt{`MW04\_no\_drhdt'}). The colourbar is set on a logarithmic scale between $1$ and $300$ clusters. Red stars correspond to the present-day masses and radii of Galactic GCs \citep{Baumgardt18}, orange diamonds indicate clusters in M31 \citep{Barmby07,Huxor14,Caldwell16}, and pink squares show young clusters in M83 \citep{Ryon15}. As clusters disrupt and lose mass, they move leftwards towards lower densities.}
    \label{fig:final_radii_mass_noevolution}
\end{figure*}

The final cluster size-mass relations for the simulations without size evolution are shown in Fig.~\ref{fig:final_radii_mass_noevolution}. As clusters evolve and disrupt, their mass decreases but their size remains unchanged. Thus, their density decreases and they move leftwards in the size-mass space relative to their initial location. The mass-loss mechanisms significantly reduce the number of surviving clusters, with the bulk of them having densities between $10$--$10^2~\msunpc$. The number of clusters with high densities ($\rho>10^4~\msunpc$) is also greatly reduced. For example, only $1.6~$per cent of clusters in the `2D distribution to $1$--$10~\myr$' population still retain those high densities.

When comparing to observational data from the Milky Way, M31 and M83 (red, orange and pink data points respectively), care must be taken as the observed clusters have limited ages while the model clusters span the whole range of ages. Nevertheless some important differences between the cluster populations arise. 
The population formed with an initial constant half-mass radius is clearly incorrect. However, its evolved cluster mass function of old clusters is in excellent agreement with that of GCs in the Milky Way and M31 \citepalias[][and see Fig.~\ref{fig:final_mass_obsdata}]{Reina-Campos22}, so some aspect underlying it makes sense. There are no surviving massive clusters ($m>10^6~\msun$) in the population formed with initial physically-motivated sizes. We also see that neither clusters in the constant density populations nor in the populations sampled from empirical fits without scatter match the observed spread in radius. And lastly, only the populations sampled from empirical fits with scatter show a range in radius comparable to the observed distributions, although there are too few low density clusters for masses above $m>10^4~\msun$. We compare our model populations and the observations at the appropriate ages in the following section.

\begin{figure*}
    \includegraphics[width=\hsize,keepaspectratio]{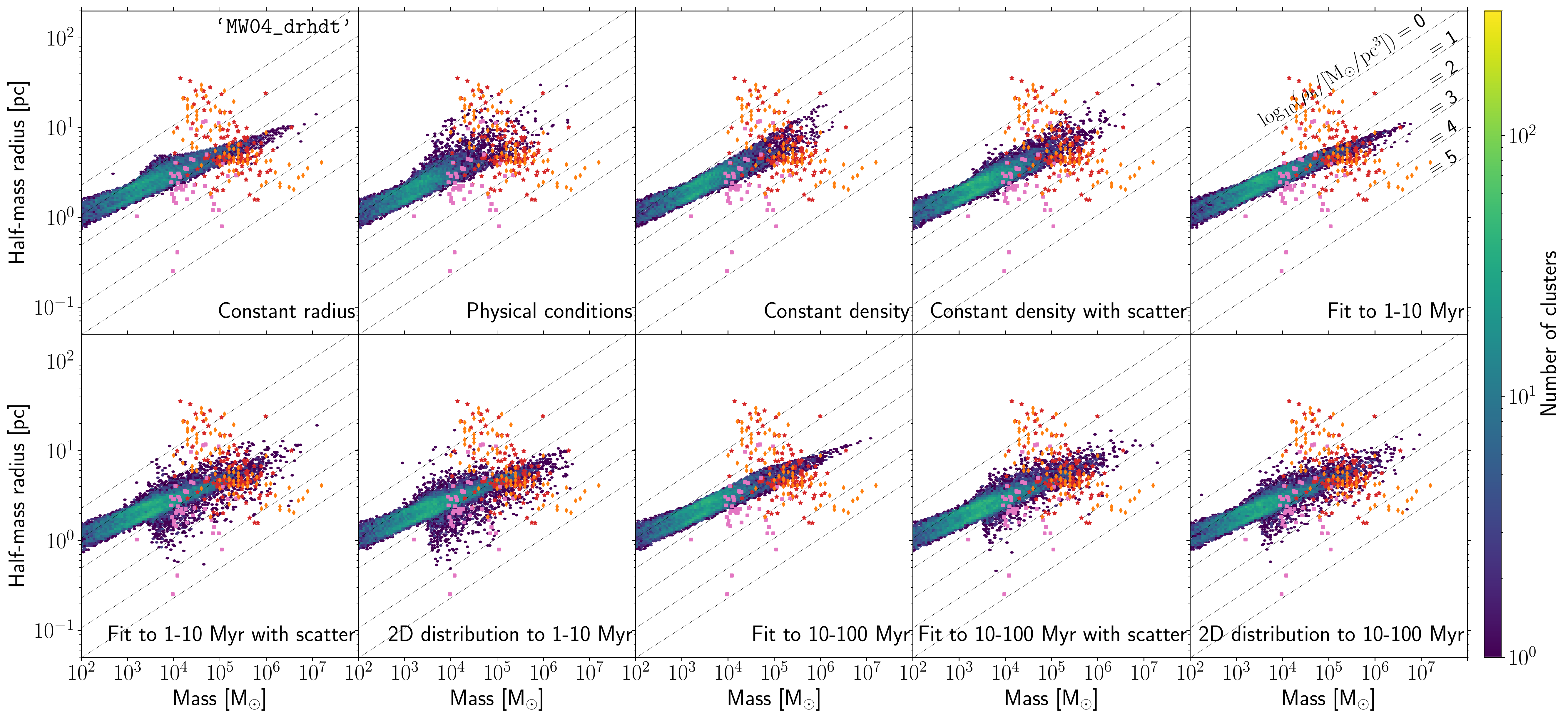}
    \caption{Cluster half-mass radius as a function of their final masses for our 10 sub-grid cluster populations in the Milky Way-mass galaxy with size evolution (\texttt{`MW04\_drhdt'}). The colourbar is set on a logarithmic scale between $1$ and $300$ clusters. Red stars correspond to the present-day masses and radii of Galactic GCs \citep{Baumgardt18}, orange diamonds indicate clusters in M31 \citep{Barmby07,Huxor14,Caldwell16}, and pink squares show young clusters in M83 \citep{Ryon15}. The size evolution model dominates over the initial assumption on the size-mass relation.} 
    \label{fig:final_radii_mass_evolution}
\end{figure*}

If the half-mass radius of the clusters is allowed to change with time, the evolved size-mass relations are strikingly similar to each other. We show the final cluster size-mass relations for the Milky Way-mass galaxy with size evolution in Fig.~\ref{fig:final_radii_mass_evolution}. Regardless of the initial size-mass model assumed, after a Hubble time of evolution all populations have evolved into a very similar shape. By fitting a linear regression model, we find that these relations have slopes of $0.19$--$0.26$, which does not correspond to the case of clusters evolving into having constant density. This indicates that the model used to describe the evolution of the sizes dominates over their initial conditions. We explore in more detail the physical mechanism responsible for this similarity in Sect.~\ref{sub:why-differences}.

In general, allowing the size to evolve does not seem to affect cluster disruption significantly. The population that is most affected by cluster disruption is the one that assumes the initial half-mass radii to be given by the natal conditions of the gas (`Physical conditions'). Only $2.2$~per cent of those clusters survive when their sizes do not evolve, and their masses represent merely $0.6$~per cent of the initial total mass in clusters. Those percentages increase to $7.1$ and $2.3$, respectively, when their sizes evolve with time. At the other extreme, clusters formed from empirical fits with scatter are the least affected by disruption, with $\sim15$~per cent of them surviving and their masses representing $\sim13$~per cent of the initial mass in clusters. These percentages decrease slightly to $\sim10$ and $\sim9$, respectively, when allowing the size to evolve. 

\subsection{Cluster property distributions }\label{sub:which-one-is-correct}

Here we examine the mass, size and density distributions of our cluster populations for different age cuts, and compare them to observed cluster masses and sizes in the Milky Way, M$31$ and M$83$.

\subsubsection{Final mass functions}

\begin{figure*}
	\includegraphics[width=\hsize,keepaspectratio]{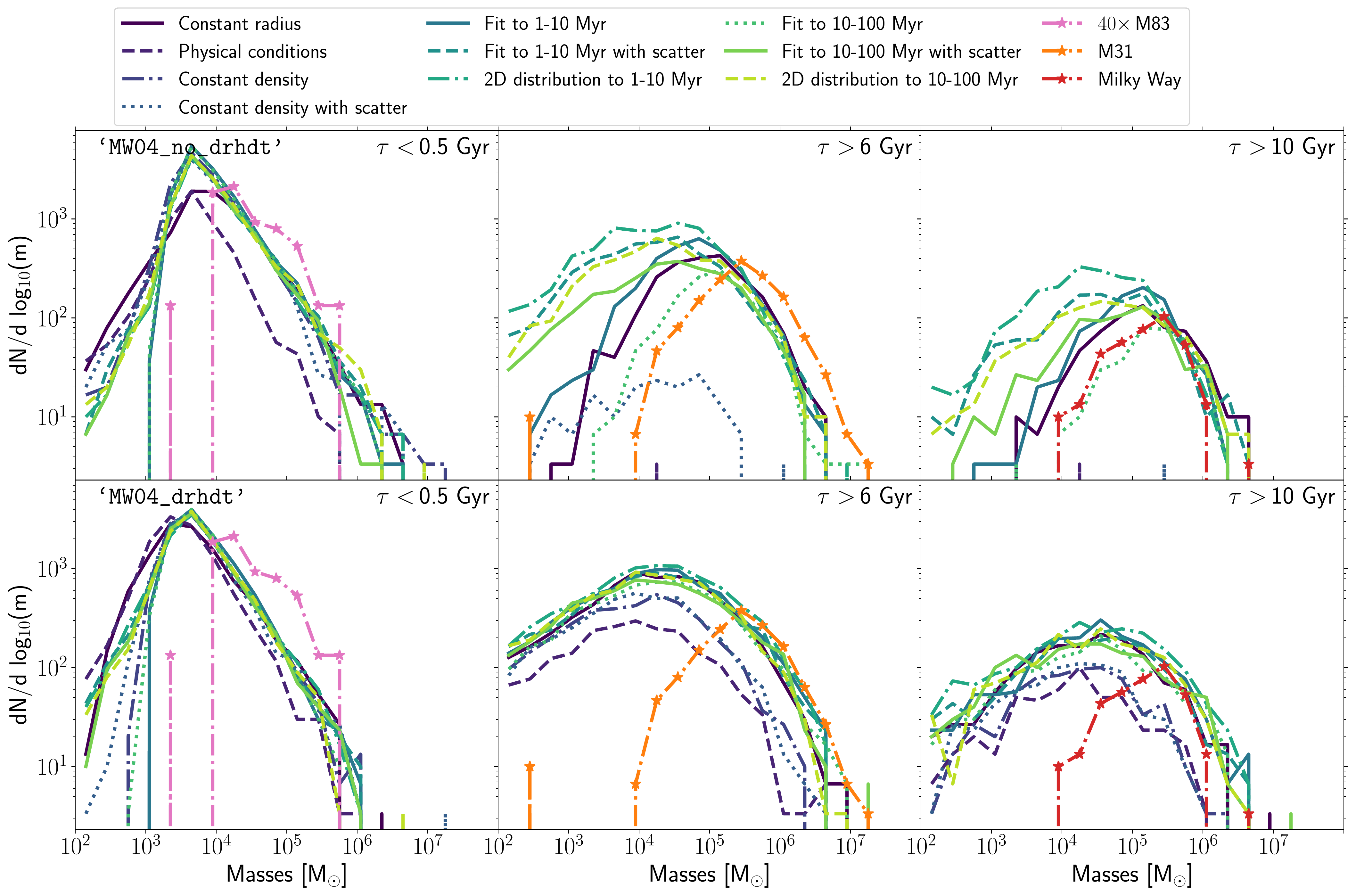}
    \caption{Comparison of the final mass functions of our 10 sub-grid cluster populations in the galaxy without (\textit{top row}, \texttt{`MW04\_no\_drhdt'}) and with size evolution (\textit{bottom row}, \texttt{`MW04\_drhdt'}). Different columns correspond to stellar clusters within a different age range, as indicated in the top-right corner of each panel. We include the cluster mass functions in the Milky Way \citep{Baumgardt18}, M31 \citep{Caldwell16} and M83 \citep{Ryon15} as dash-dotted lines with stellar symbols. 
    }
    \label{fig:final_mass_obsdata}
\end{figure*}

The mass distributions of our sub-grid star cluster populations are shown in Fig.~\ref{fig:final_mass_obsdata} for three different age cuts, chosen so that we can compare them to the observed masses of clusters in the Milky Way \citep{Baumgardt18}, M31 \citep{Caldwell16} and M83 \citep{Ryon15}. Overall, we find that around $90~$per cent of clusters are completely disrupted, and only a small subset survives to old ages. In all our simulations, only the most massive clusters survive for long times. To put it another way, the peak of the mass function moves towards higher masses with time, due to disruption, from $\sim 10^4~\msun$ for young clusters to $\sim10^5~\msun$ for old clusters. Disrupted clusters tend to have initial masses below $m<10^5~\msun$. Regardless of the assumption of the initial size-mass relation, young star clusters ($\tau < 0.5~\gyr$) exhibit a power-law mass function with slope similar to that in M83. This power-law shape is given by the initial cluster mass function, and indicates than within the first $500~\myr$ disruption has not significantly modified the cluster mass function.

Clusters of intermediate ages ($\tau > 6~\gyr$) have masses distributed differently based on the different initial size-mass relations, and on whether the size is allowed to evolve. If the size is kept constant (upper middle panel in Fig.~\ref{fig:final_mass_obsdata}), clusters simulated either with the empirical fits without scatter or with a constant radius are the best match to the observed mass function in M31. Those populations show log-normal distributions with peaks at $M\sim10^5~\msun$. In contrast, when scatter is considered together with the empirical fits, there are too many low-mass clusters and the peak shifts to $M\sim10^4~\msun$. Lastly, clusters simulated with an assumed constant initial density with scatter are significantly disrupted and their mass function is flat with much lower normalization. If instead the size of clusters is allowed to evolve (lower middle panel in Fig.~\ref{fig:final_mass_obsdata}), the cluster mass distributions are very similar between all our models, and are well represented by a log-normal with a peak at $M\sim5\times10^{4}~\msun$, with a high-mass distribution very similar to that of M31. The similarity between models indicates that the size evolution prescription homogenizes the cluster populations. 

The mass distributions of old star clusters ($\tau>10~\gyr$) are also well described by log-normal functions, with the details depending on the initial size-mass relation and evolution model considered. When the size is kept constant (upper right panel in Fig.~\ref{fig:final_mass_obsdata}), the observed mass function in the Milky Way is best-matched by the populations modelled with a constant radius and the empirical fits without scatter. If scatter is included in the initial empirical fit, too many low-mass clusters survive to these old ages. This implies that disruption is not sufficiently effective at these low masses, probably due to their high densities. 

In contrast, there are no surviving intermediate-age clusters when assuming an initial constant density or physically-motivated initial size. And this extends to the population modelled with an initial constant density with scatter when examining old clusters. These initial size-mass relations thus fail at reproducing the observed mass functions in M31 and the Milky Way. These populations have few clusters with initial densities higher than $\rho>10^3~\msunpc$, which suggests that a high initial density is a pre-requisite for cluster survival over a Hubble time. We explore the physical drivers behind these distributions in Sect.~\ref{sub:why-differences}. 

\subsubsection{Final radii distributions}

\begin{figure*}
	\includegraphics[width=\hsize,keepaspectratio]{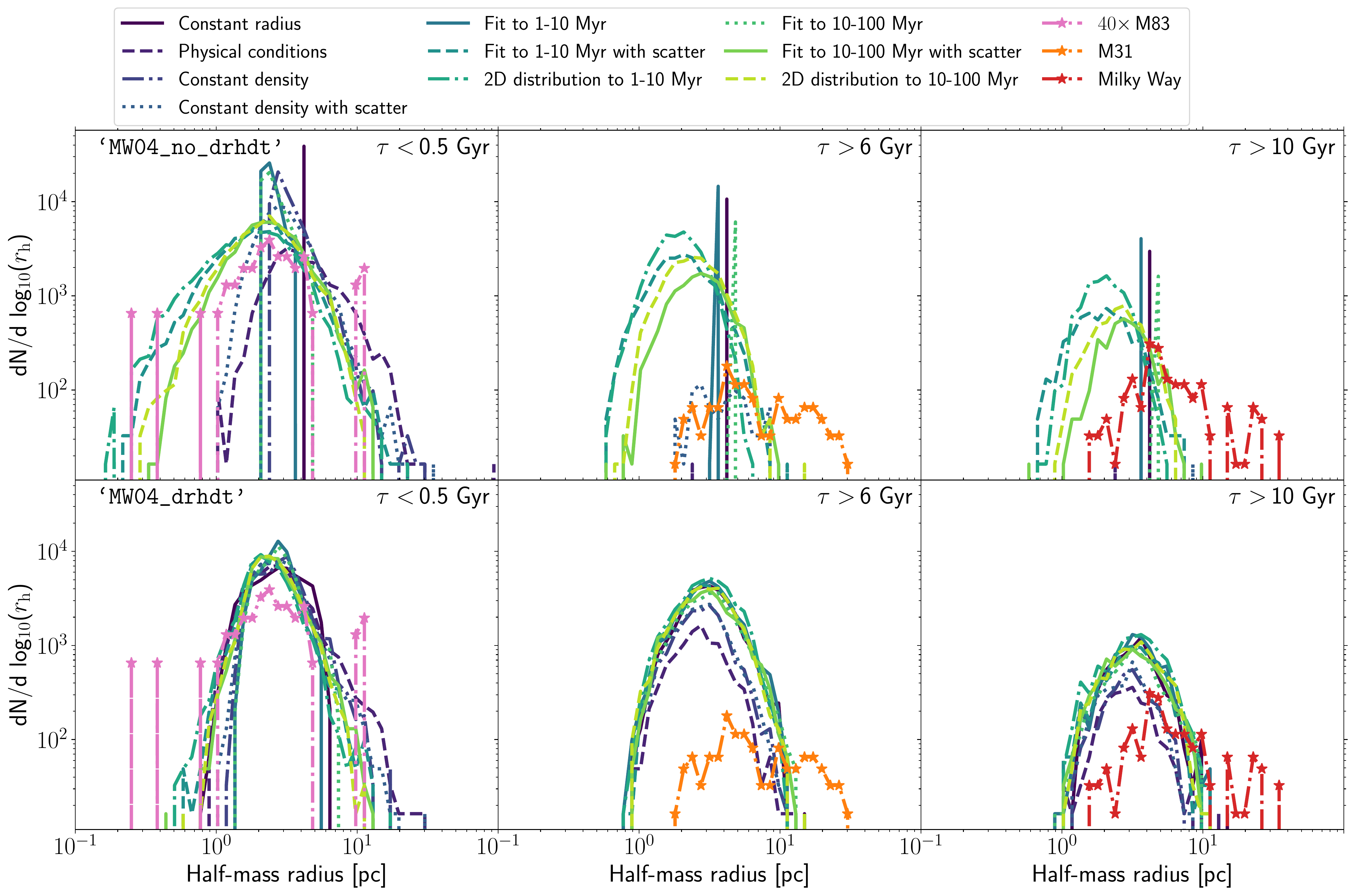}
    \caption{Comparison of the final radius functions of our 10 sub-grid cluster populations in the galaxy without (\textit{top row}, \texttt{`MW04\_no\_drhdt'}) and with size evolution (\textit{bottom row}, \texttt{`MW04\_drhdt'}). Different columns correspond to stellar clusters within a different age range, as indicated in the top-right corner of each panel. We include the cluster radius distributions in the Milky Way \citep{Baumgardt18}, M31 \citep{Barmby07,Huxor14,Caldwell16}, and M83 \citep{Ryon15} as dash-dotted lines with stellar symbols.}
    \label{fig:final_radius_obsdata}
\end{figure*}

We now turn to examine the final radii distributions of our simulated cluster populations, shown in Fig.~\ref{fig:final_radius_obsdata}. Young clusters ($\tau<0.5~\gyr$) without size evolution show a variety of distributions: those formed with an initial linear fit exhibit a narrow range in their sizes, whereas those formed from empirical fits with scatter over-predict the number of clusters with small ($\rh<1~\pc$) and large ($\rh>6~\pc$) sizes. The peaks of the distributions fall between $\sim2$--$4~\pc$, in rough agreement with the population of young clusters in M83. In contrast, when the size is allowed to evolve, the populations are already  homogeneous within $500~\myr$ and all the distributions peak at $\sim2.5~\pc$. 

Older clusters fail to reproduce the range in sizes observed in M31 and in the Milky Way. When the size is constant, our models fall in two categories. For clusters formed with empirical fits with scatter, there are too many surviving small ($\rh<2~\pc$) ones. This suggests that these models over-predict the number of compact and dense clusters, which are more resilient against tidal shock disruption and thus more likely to survive. Secondly, for clusters that formed from empirical linear fits, only those that were initially more massive than $10^5~\msun$ and had initial constant sizes survive to ages older than $10~\gyr$. Their lack of a size spread is in stark contrast with the range shown by the GC population in M31 and in the Milky Way. 

When the size is allowed to evolve, all populations exhibit a log-normal distribution peaked at $~3.5~\pc$ that does not extend past $10~\pc$. None of our models reproduce the more extended clusters in the Milky Way. These are NGC2419 (which is a loose and distant cluster likely associated with the Sagittarius dwarf galaxy, \citealt{Massari2019}), Palomar 3, 5, 13, and 15, and Sagittarius II. Palomar 5 shows very clear tidal tails signaling that is close to dissolution \citep{Bonaca20}, and there is evidence for extra-tidal structure around Palomar 3 \citep{Sohn2003}, Palomar 15 \citep{Myeong2017} and NGC 2419 \citep{Kundu2022}. Our size evolution model does not encapsulate the final stages of cluster disruption, so we may not be able to correctly capture these very extended clusters. 

\subsubsection{Final density distributions}

\begin{figure*}
	\includegraphics[width=\hsize,keepaspectratio]{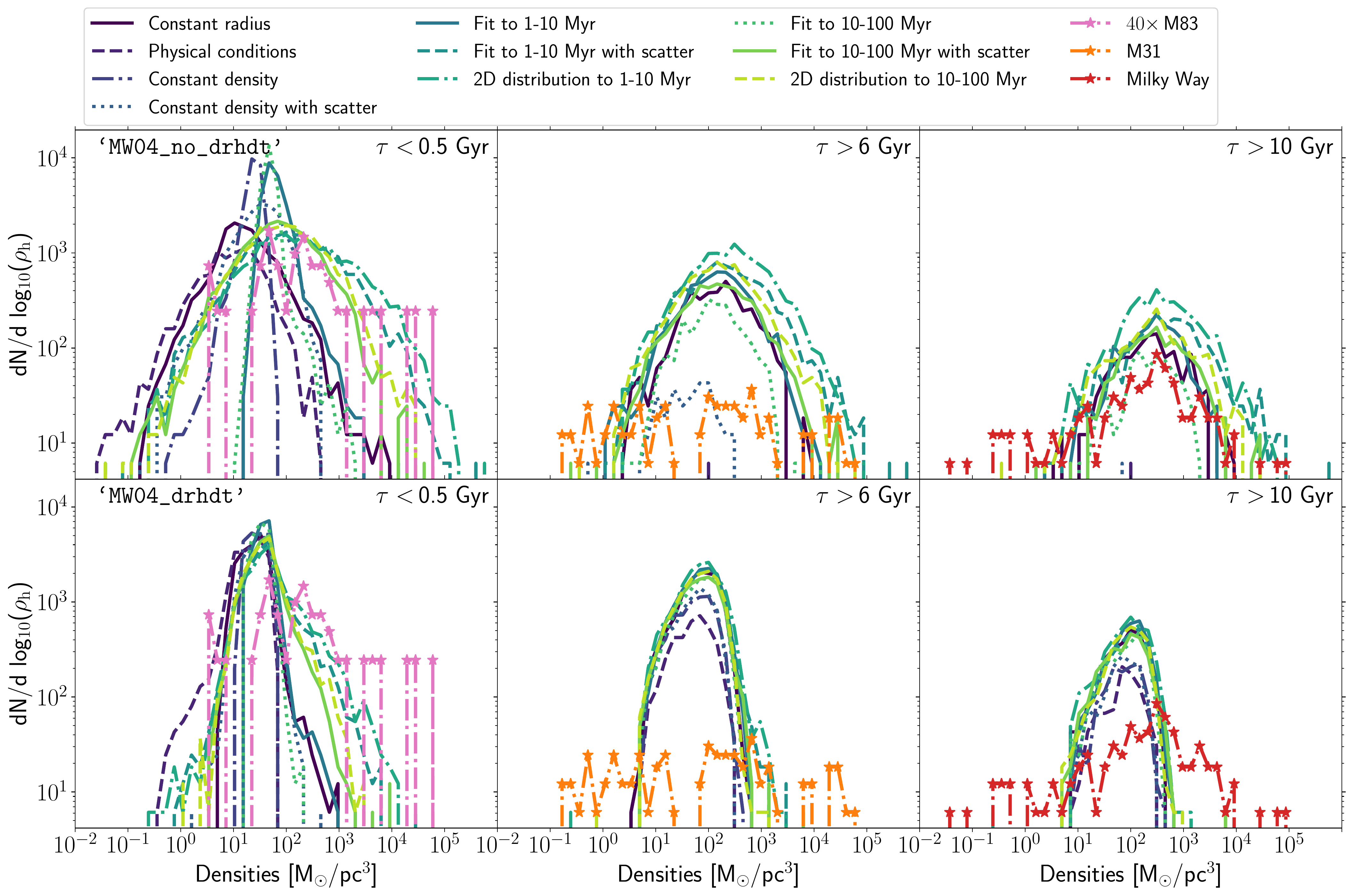}
    \caption{Comparison of the final density functions of our 10 sub-grid cluster populations in the galaxy without (\textit{top row}, \texttt{`MW04\_no\_drhdt'}) and with size evolution (\textit{bottom row}, \texttt{`MW04\_drhdt'}). Different columns correspond to stellar clusters within a different age range, as indicated in the top-right corner of each panel. We include the cluster mass functions in the Milky Way \citep{Baumgardt18}, M31 \citep{Barmby07,Huxor14,Caldwell16}, and M83 \citep{Ryon15} as dash-dotted lines with stellar symbols.}
    \label{fig:final_density_obsdata}
\end{figure*}

Fig.~\ref{fig:final_density_obsdata} shows the distributions of densities at the half-mass radius for our cluster populations. As we saw in the case of the masses and the radii, the distributions of densities of young clusters without size evolution are the most diverse. Overall, our cluster populations extend to densities much lower ($\rho<3~\msunpc$) and higher ($\rho>10^4~\msunpc$) than the majority of those observed in M83. Clusters that form with a constant radius or density, or from a physically-motivated size tend to have lower densities with a peak near $\sim10~\msunpc$, whereas those that form from empirical fits with scatter exhibit higher densities with peaks at $\sim100~\msunpc$. When the size evolves, the range in densities decreases and becomes similar to that in M83, although clusters are in general too sparse compared to the observed objects (peaks at $\sim 30~\msunpc$).

Despite the differences in mass and radius, the density distributions of old clusters ($\tau>10~\gyr$) without size evolution are in good agreement to the one observed in the Milky Way. Our simulated density distributions exhibit a log-normal function with a peak at $300~\msunpc$. Regardless of this good agreement, those cluster populations formed from empirical fits with scatter tend to have an excess of high-density clusters. In contrast, when the size is allowed to evolve, the distributions peak at lower densities, $\sim100~\msunpc$, and are too narrow in comparison to the one observed in the Milky Way.

In summary, we find that none of our models fully reproduces the mass, radius and density distributions of M83, M31 and the Milky Way. The populations that best match the mass and density distributions are those that formed either with an initial constant radius, or from empirical linear fits without scatter. However, their lack of radii spread is in disagreement with the observed size distributions. When scatter is introduced in the empirical fits, initially extremely dense ($\rho>10^4~\msunpc$) and low-mass clusters survive to ages older than $10~\gyr$. Their high densities make them very resilient against tidal shock disruption, thus sheltering them from complete dissolution over cosmic time. 

\subsection{How does cluster size distribution and evolution affect cluster disruption processes?}\label{sub:why-differences}

In this section we explore the physical mechanisms that determine the evolution of our cluster populations. To simplify our discussion, we restrict our analysis to five representative cluster populations out of the ten initial size-mass relations considered. We consider our two simplistic assumptions (constant radius and constant density), the population where the cluster sizes depend on the physical conditions of the gas, and two versions of the empirical fits, one with and one without scatter.

In our simulations, our clusters lose mass due to stellar evolution, tidal shocks, evaporation of stars due to two-body relaxation, and dynamical friction (see Sect.~\ref{sec:methods}). Disruption driven by tidal shocks and evaporation  strongly depend on both the mass and the size of the cluster, whereas stellar evolution results in a constant fractional mass loss from each cluster, and dynamical friction depends more strongly on the cluster's position in the galaxy. Since these last two properties will be the same across all our models, we can use our results to investigate the relative importance of shocks and evaporation on cluster evolution, for different assumptions about the initial size distributions of the clusters. 

\begin{figure*}
    \includegraphics[width=\hsize,keepaspectratio]{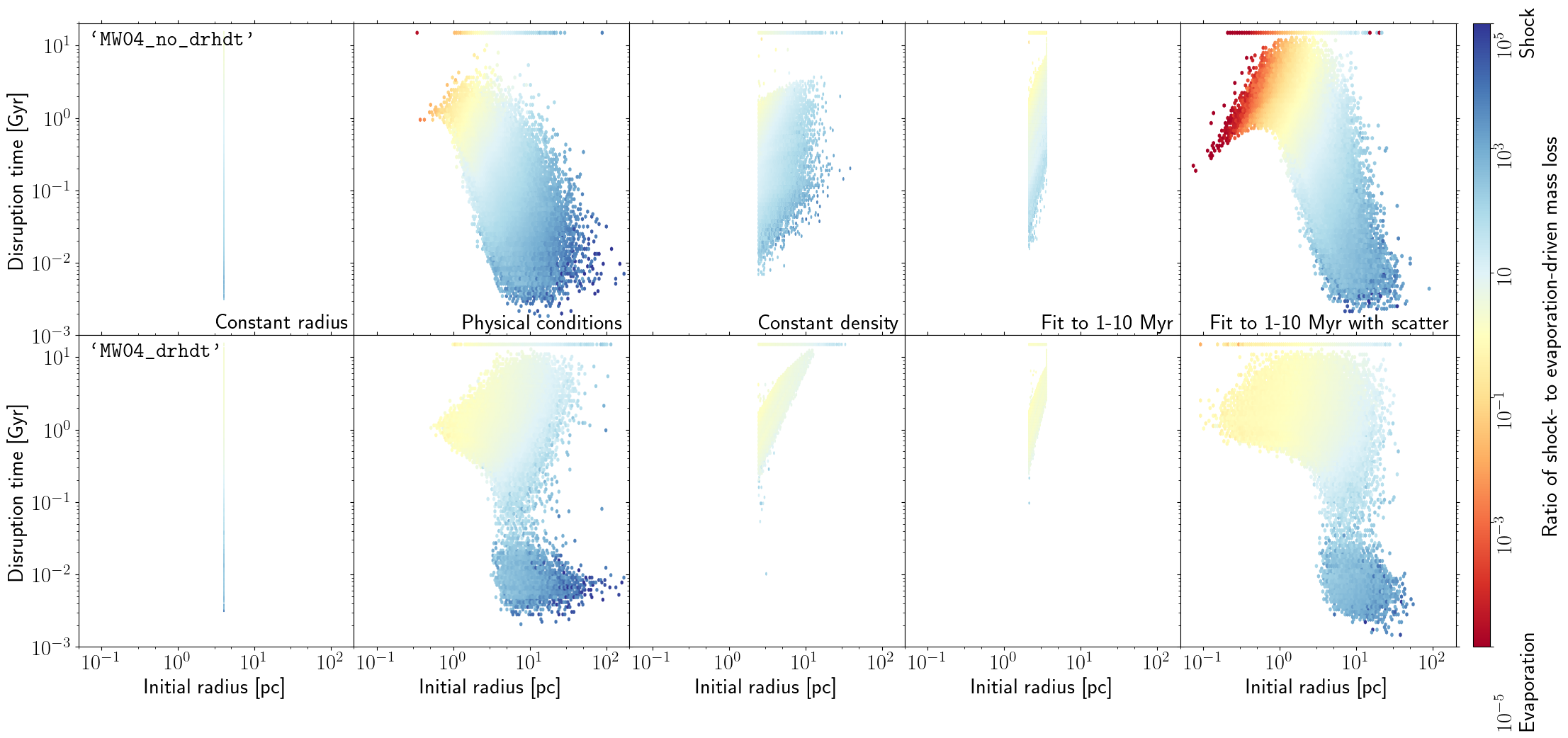}
    \caption{Disruption timescales as a function of the initial half-mass radius for a subset of 5 cluster populations in the Milky Way-mass galaxy without (\texttt{`MW04\_no\_drhdt'}, \textit{top row}) and with size evolution (\texttt{`MW04\_drhdt'}, \textit{bottom row}). Surviving clusters are given a disruption timescale of $15~\gyr$. The colourbar corresponds to the ratio of the mass lost due to tidal shocks over evaporation, with red and blue indicating that evaporation or tidal shocks dominate, respectively. Extended clusters ($\rh>5~\pc$) are quickly disrupted by tidal shocks due to their low densities, whereas smaller clusters are mostly disrupted due to evaporation over longer timescales ($t_{\rm dis} \sim 1$--$10~\gyr$). }
    \label{fig:disruption_time_radii}
\end{figure*}

The times to disruption for clusters as a function of their initial radius is shown in Fig.~\ref{fig:disruption_time_radii}. Each cluster is also colour-coded by their dominant disruption mechanism. We quantify this by using the ratio of the amount of mass lost from the cluster due to shocks over the amount lost due to evaporation. Dark blue points are shock-dominated and dark red points are evaporation-dominated.

The top row shows the models where the cluster radius does not evolve with time. Extended clusters ($\rh>10~\pc$) are very susceptible to shocks, and strong shocks can destroy a cluster very quickly ($t_{\rm dis}\sim5~\myr$--$0.5~\gyr$). Since the clusters are not permitted to shrink due to mass loss, they stay large and continue to be strongly affected by shocks. Compact clusters ($\rh<1~\pc$) are protected against shocks and so are mostly disrupted by evaporation. Since the clusters stay compact even when they lose mass, they continue to be evaporation-dominated. Two-body relaxation occurs on the cluster's internal dynamical time, and is a slower process than shocks, so evaporation becomes important after $t_{\rm dis}\sim1~\gyr$ for all but the very smallest clusters (which have the shortest relaxation times). Clusters that survive to the present day (horizontal line at the top of each plot) show the same behaviour as their fully disrupted counterparts: small clusters primarily lose mass due to evaporation while large clusters are dominated by shocks. The two models with the largest range of cluster radii (``Physical conditions'' and ``Fit with scatter'') show this dichotomy most clearly, while the other models map out the same behaviour in parameter space for their more limited range of initial radii. The main reason that the ``Physical conditions'' model loses most of its clusters is that the radius distribution is skewed to high values, and shocks immediately destroy most of these large clusters. 

If size evolution is included in our models (bottom row of Figure~\ref{fig:disruption_time_radii}) then the clusters adjust as they lose mass. The largest clusters that are subjected to the strongest shocks are destroyed almost immediately (the concentration of dark blue points in the bottom right of the panels). However, if the shock is not quite so strong, the cluster has time to shrink. Its radius is now closer to its tidal radius, and shocks are not as effective at removing mass. This is seen as a lack of clusters with disruption times around $t_{\rm dis}\sim50~\myr$ compared to the upper panels. Evaporation becomes more important in the disruption of clusters with radii around 10 pc, as seen by the increase in the yellow colour (indicating approximately equal contributions of shocks and evaporation) of these clusters. Compact clusters expand as they lose mass due to evaporation, and so shocks can now contribute more to their disruption, also pushing more clusters towards yellow colours. For similar reasons, the timescale for cluster dissolution generally becomes longer for the population. With fewer clusters being primarily destroyed by (quick) shocks, the longer-timescale evaporation process contributes more and therefore more clusters can survive in all models. 

\begin{figure*}
    \includegraphics[width=\hsize,keepaspectratio]{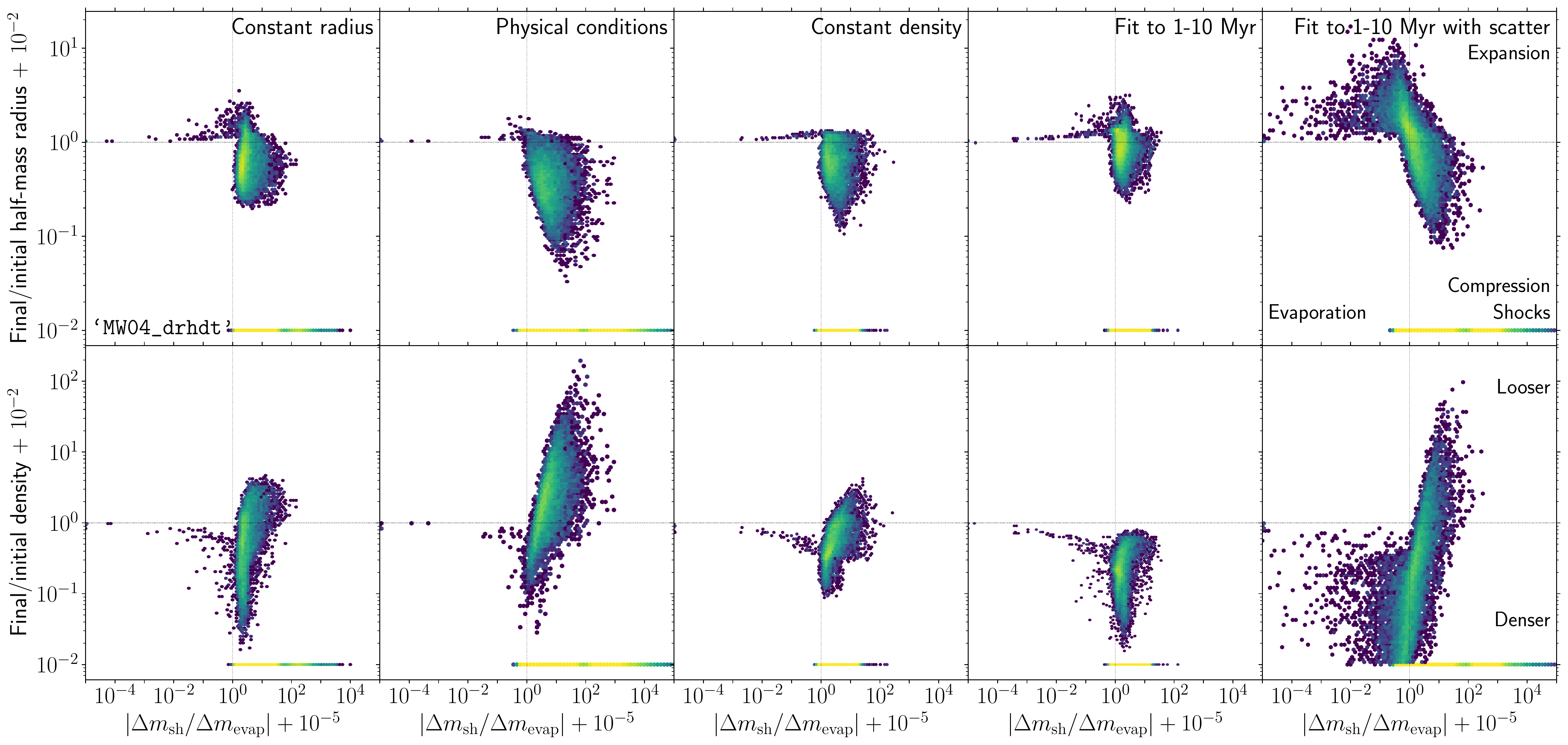}
    \caption{Ratio of the final-to-initial radius (\textit{top row}) and density (\textit{bottom row}) as a function of the ratio of shock-to-evaporation driven mass loss for a subset of our cluster populations in the Milky Way-mass galaxy with size evolution (\texttt{`MW04\_drhdt'}). The colourbar is set on a logarithmic scale between $1$ and $300$ clusters. Clusters that are completely disrupted are shown with radius and density ratios of $10^{-2}$. Initially extended clusters contract due to compressive shocks, allowing them to become resilient against further shock disruption.}
    \label{fig:ratio_radii_density_evolution}
\end{figure*}

We concentrate now on the surviving clusters, and present their change in radius (top row) or density (bottom row) as a function of the ratio of shock to evaporation-driven mass loss in Fig.~\ref{fig:ratio_radii_density_evolution}. Clusters that were originally very extended, for example those formed under the assumption of a physically motivated size, are strongly affected by shock disruption and contract. Their density increase of close to two orders of magnitude implies that these objects become very resilient to shock disruption, thus allowing them to survive to old ages ($\tau>10~\gyr$). Similarly, extended clusters formed from empirical fits with scatter also contract and increase their densities. In addition, the compact objects ($\rh<2~\pc$) that form under this assumption increase their radii and become less dense due to the effect of two-body relaxation. This transformation makes clusters more susceptible to being disrupted by shocks, and there are very few compact surviving clusters older than $500~\myr$.

In essence, the effect of radius evolution in our models is to allow clusters to move to ``safe'' regions of parameter space. Clusters that are larger than their tidal radii lose mass due to shocks and subsequently shrink; clusters that are very compact lose mass due to evaporation and then expand. Shocks are faster than evaporation, but both processes act on timescales shorter than a Hubble time, so that by the end of our simulations, the surviving clusters are only those who have the sizes and densities where neither shocks nor evaporation are particularly destructive. This is why all the final size-mass distributions from Figure \ref{fig:final_radii_mass_evolution} are very similar. A similar result was found in \citet{Gieles16}, who showed that a balance between relaxation-dominated evolution and encounter-dominated evolution creates an equilibrium density that is proportional to cluster mass to the 2/3 power, or radius goes as mass to the power of 1/9. As a reminder, our equilibrium results are more like $R \propto M^{0.2}$ (see Section \ref{sec:results}). Although our size evolution prescription is based on that of \citet{Gieles16}, there are a few differences. They determine their equilibrium configuration to be where the timescales of shock-driven and evaporation-driven mass loss are equal. Because our clusters can move around their host galaxy, and because the galaxy itself is growing and merging, the timescale for shock-driven mass loss is a complicated function of time, galactic environment, and the local environment of the cluster. We have also run simulations in which we allow the sizes to evolve only due to each of the terms in equation (\ref{eq:size-evolution-complete}). With these tests, we have confirmed that shock-driven evolution dominates. In our models, shocks are responsible for more mass loss than evaporation for most clusters. Therefore, our ``safe'' region of parameter space is not quite the same as the attractor solution from \citet{Gieles16}, but shares many of the same underlying physical motivations. 

\section{Summary and Discussion}\label{sec:conclusions}

In this paper we examine the influence of the initial cluster size-mass relation on the demographics of star cluster populations over a Hubble time. We test ten different initial size-mass relations used in the literature, and evolve a Milky Way-mass galaxy with ten parallel sub-grid cluster populations using the \emppathfinder framework. We keep all the other input models the same across the populations, so the only difference is on their initial sizes. We repeat the cosmological zoom-in simulation, this time allowing the initial size to evolve over time, and we characterize their final size-mass relations (Sect.~\ref{sub:final-size-mass-relations}). 

We find that none of our models simultaneously reproduce both the cluster mass function and cluster radius distribution after 6-10 $\gyr$ of evolution. We can successfully reproduce the cluster mass function for old clusters with models that have a small range of initial radii (constant radius or fits to observations without scatter), and do not allow the radii to evolve with time. However, these models do not agree with the observed radius distribution of clusters -- real cluster populations have a range of radii at all ages. 

All our models with size evolution result in approximately the same distribution of radii at old ages, but contain too many low-mass clusters. Physically, we expect that clusters should change their sizes with time, as that behaviour is seen in all detailed dynamical evolution of clusters in any galactic potential. Requiring clusters to keep a constant radius even as they lose mass is unphysical. We note that the peak of the radius function in our models that allow size evolution is approximately 3--4 pc for our oldest clusters, which is not unreasonable for the Milky Way population. We suggest that the largest clusters currently seen in the Milky Way are not produced by our models in part because we do not accurately model the final stages of tidal disruption. 

There are a few choices for initial clusters sizes that are entirely ruled out by our models. Cluster populations that form with an initial constant density (`Constant density' and `Constant density with scatter'), or with initial sizes based on the physical conditions of their natal sites (`Physical conditions') do not reproduce any of the observed mass, radius or density distributions. These populations form very extended clusters that disrupt quickly due to tidal shocks, so there are no intermediate nor old clusters unless the clusters are allowed to change their size. Even with size evolution, these final populations have fewer clusters, and almost no massive clusters, in contradiction to the observations.

We find that there is a region of parameter space in cluster radius where clusters are optimally protected from both shocks and evaporation. Typically, large clusters are destroyed by shocks, while small clusters are destroyed by evaporation. Size evolution allows clusters to move into this protected region, because large clusters are allowed to shrink, while small clusters can expand. In both cases, the clusters become less affected by their primary mass loss mechanism, thus allowing them to survive to the present day. 

We conclude that our size evolution prescription, while helpful for allowing us to explore physical processes of cluster dissolution, is too simplified to reproduce the observed radius and mass distributions of real clusters. Our prescription is a semi-analytic description of cluster evolution on circular orbits in the Milky Way potential, while studies of cluster evolution in eccentric and inclined orbits show complicated, orbit-dependent behaviour \citep{webb2014}. Our prescription also assumes fully adiabatic expansion after mass loss, and does not include effects such as the response time of the cluster if its tidal radius is changing quickly. 
We also note that the mass loss prescriptions that are used in this work and similar implementations have been tested under only a few different choices for cluster radius, and have not typically been tested in models where the radius is allowed to change with time. Improved understanding of both the mass and radius evolution of clusters in realistic, time-varying potentials, would be helpful in resolving some of our discrepancies. 

Young massive clusters are observed to have a large scatter in radius, even at ages less than 10 $\myr$. The results of these simulations suggest that the vast majority of these young clusters will not survive to very old ages, and that the largest clusters will be fully disrupted very quickly. The location of these clusters at the moment of dissolution will determine whether their stars go on to form the halo or the disk of the galaxy. 

Semi-analytic models of cluster evolution will be used as sub-grid prescriptions in galaxy evolution models for the foreseeable future. Therefore, it is important that we make those models as accurate as possible in order to be able to make appropriate inferences about the interplay between galaxy evolution and cluster properties. This work concentrated on the treatment of cluster radii, and demonstrated that our current approach is not yet sufficient. We make the following suggestions for improvement of the models. First, it is important that the underlying approach to tidal effects on clusters is handled as accurately as possible. The spatial resolution of the simulation must be carefully considered, as must the treatment of compressive vs expansive tides, impulsive shocks compared to adiabatic damping, and the orientation of the tidal field and its evolution with time.

In the context of the commonly-used tidal tensor approach to tidal heating used in this and other models, size evolution of clusters should not be ignored. Because size evolution allows clusters to move to the ``safe" regions of parameter space and therefore essentially erases the initial mass-radius distribution, treating size evolution correctly is more important than determining the correct initial mass-radius distribution. At the moment, the reaction of cluster radii to tidal shocks is parameterized with a single parameter that includes an assumption about the density profile and other structural properties of the cluster (e.g. isotropic and in equilibrium), and the impulsive nature of the tidal event. That is a lot of work for a single parameter to do, especially because shocks dominate the mass and radius evolution of clusters. There are detailed N-body models of shocks which can be used to either test or improve the simple parameterization \citep[e.g.][]{MartinezMedina2022}. That paper concluded that the density structure of the cluster was important. Future work could include that effect. We expect that would help explain the large radius/low density clusters that are not seen in our current models.

The size evolution of clusters due to two-body relaxation is assumed in these simple models to be both instantaneous and as a result of a constant fraction of stars being lost per relaxation time. We know, however, that for clusters on orbits that are not circular and in the disk of the Milky Way, neither of these assumptions are correct. Clusters on eccentric orbits can lose stars at perigalacticon and then recapture them later in the orbit \citep{webb2014a}. We also know that the fraction of stars that are lost depend on the inclination and eccentricity of the orbit \citep{webb2014} and where the cluster is along that orbit. A more complex treatment of size evolution due to relaxation is also required.

Including more correct but more complex treatments of mass and radius evolution due to both shocks and relaxation will require appropriate understanding of the distribution of cluster orbits in a forming galaxy, the expected distribution of cluster density profiles, as well as other properties. These will necessarily add potential uncertainty into our understanding of cluster evolution in the context of galaxy formation, but it also holds the promise of increased accuracy of our models, allow for more detailed comparison with observations, and hence improved understanding. Steps along this path must be taken with care and rigor.

\section*{Acknowledgements}
MRC thanks the participants of the Aspen Workshop "Illuminating Galaxy Formation with Ancient Globular Star Clusters and Their Progenitors" for discussions that led to this project. 
MRC gratefully acknowledges the Canadian Institute for Theoretical Astrophysics (CITA) National Fellowship for partial support; this work was supported by the Natural Sciences and Engineering Research Council of Canada (NSERC). AS is supported by the Natural Sciences and Engineering Research Council of Canada. GB acknowledges financial support from the MITACS Globalink Fellowship program.

\textit{Software}: This work made use of the following \code{Python} packages: \code{h5py} \citep{h5py_allversions}, \code{Jupyter Notebooks} \citep{Kluyver16}, \code{Numpy} \citep{Harris20}, \code{Pandas} \citep{McKinney10, pandas_allversions}, and \code{Scipy} \citep{Jones01}, and all figures have been produced with the library \code{Matplotlib} \citep{Hunter07}. The comparison to observational data was done more reliably with the help of the \code{webplotdigitizer}\footnote{https://apps.automeris.io/wpd/} webtool.

%%%%%%%%%%%%%%%%%%%%%%%%%%%%%%%%%%%%%%%%%%%%%%%%%%
\section*{Data Availability}

The data underlying this article will be shared on reasonable request to the corresponding author.

%%%%%%%%%%%%%%%%%%%% REFERENCES %%%%%%%%%%%%%%%%%%

% The best way to enter references is to use BibTeX:

\bibliographystyle{mnras}
\bibliography{Sizes} % if your bibtex file is called example.bib

% Don't change these lines
\bsp	% typesetting comment
\label{lastpage}
\end{document}